\documentclass[12pt]{article}
\usepackage{graphics}
\renewcommand{\theequation}{\arabic{section}.\arabic{equation}}
\begin{document}
%------------------------------------------------------------------------------
\markright{Gravitons ...}
%------------------------------------------------------------------------------
\title{Can Gravitons Be Detected?}

\author{Tony Rothman ~$^*$ and Stephen Boughn$^\dagger$
\\[2mm]
{\small\it \thanks{trothman@princeton.edu}}~ \it Princeton
University,
\\ \it Princeton, NJ 08544
\\
{\small\it\thanks{sboughn@haverford.edu}}~ \it Haverford College, \\
\it Haverford, PA 19041 }

\date{{\small   \LaTeX-ed \today}}
%-----------------------------------------------------------------------------

\maketitle
%-----------------------------------------------------------------------------

%-----------------------------------------------------------------------------
\begin{abstract}
Freeman Dyson has questioned whether any conceivable experiment in
the real universe can detect a single graviton.  If not, is it
meaningful to talk about gravitons as physical entities? We
attempt to answer Dyson's question and find it is possible concoct
an idealized thought experiment capable of detecting one graviton;
however, when anything remotely resembling realistic physics is
taken into account, detection becomes impossible, indicating that
Dyson's conjecture is very likely true. We also point out several
mistakes in the literature dealing with graviton detection and
production.

\vspace*{5mm} \noindent PACS: .03.65.Sq, .04.30.Db, .04.30.Nk,
.04.80.Cc, .04.80.Nn, 95.30.Cq
\\ Keywords: Gravitons, Gravitational Waves, Gravitational Bremsstrahlung,
Graviton-electron Cross Sections.

\end{abstract}
%-----------------------------------------------------------------------------
\section{Introduction}
\label{intro}
%-----------------------------------------------------------------------------
The search for gravitational waves, one of the central predictions
of general relativity, has been ongoing for several decades.
Strong indirect evidence for their existence comes from the timing
of the orbital decay rate of the binary pulsar PSR1913+16, and
with the completion of LIGO II and other gravity-wave
observatories, researchers expect a direct detection. Gravitons
themselves present a thornier issue. Although physicists routinely
speak as if bosons mediating the gravitational force exist, the
extraordinary weakness of the gravitational interaction makes the
detection of one gravitational quantum a remote proposition.
Recently, Dyson\cite{Dyson04} has suggested that detection of a
single graviton may in fact be ruled out in the real universe. If
so, is it meaningful to talk about gravitons as physical, or do
they become metaphysical entities?

In this note we attempt to answer Dyson's question ``in
principle."   For both physical and philosophical reasons the
matter turns out to be not entirely trivial, and both
considerations require that the rules of the game be defined at
the outset. We concede at once that there appear to be no
fundamental laws disallowing the detection of a graviton, and so
we take the approach of designing thought experiments that might
be able to detect one.  Such an ``experimental" approach, however,
also has ambiguities; allowing an infinite amount of time for an
experiment in any imaginable universe would certainly allow the
detection of anything. Thus we impose a few physical restrictions:
We consider only experiments using standard physics in four
dimensions that could be performed in something like the age of
the real universe.

One must also agree on what sort of experimental result one would
accept as establishing the existence of a quantum of gravity. When
one recalls that to establish the photon picture of light took
 a decade, if not two, one sees that this ``philosophical" issue
 requires perhaps even more negotiation than the physical one.
We discuss the matter in greater detail in \S \ref{heuristic} and
\S \ref{discussion}. Moreover, in addressing Dyson's conjecture,
one is inevitably led to rather subtle physics, which apart from
its own intrinsic interest, has caused some confusion and errors
in the literature. For pedagogical reasons, we attempt to explain
the basis for our calculations as well as the others, leaving
however most of the fine details to the Appendix and a more
technical companion paper\cite{BR05}, henceforth BR.

Should Dyson's question be answered in the affirmative, that is,
if one decides that detection of a single graviton is physically
impossible, this immediately raises the issue of whether it is
necessary to quantize gravity.  There are of course theoretical
reasons for wanting to do so---after all, every other fundamental
field is quantized---and there has been some discussion in the
literature about the consistency of gravitational theories without
quantization \cite{Ros63,EH77,PG81,Sm85}.  These discussions
appear to be inconclusive and we do not enter into them.  Rather,
we restrict ourselves to the more limited question we have already
set forth: can one experimentally detect a graviton?

To begin, we assume the most general physics and units in which $c
= 1$, but $G=G$ and $\hbar = \hbar$, and write down the criterion
to detect a graviton.

%-----------------------------------------------------------------------------
\section{Detection Criterion}
\setcounter{equation}{0} \label{detection}
%-----------------------------------------------------------------------------
Regardless of the exact nature of the detector or interaction, we
demand that to detect a single graviton with high probability, its
path-length $\lambda$ in the detector  should exceed a mean-free-
path, or
\begin{equation}
n\sigma\lambda\ge 1 \; , \label{sdc}
\end{equation}
where $\sigma$ is the interaction cross section and $n$ is the
density of detector particles.  We wish to maximize the left hand
side of this expression. For simplicity we consider a detector of
atomic hydrogen, which implies $n = M_d/(m_pV)$, with $M_d$ the
detector mass, $V$ its
 volume and and $m_p$ the proton mass. Assume the detector is spherical
 with a diameter $\ell = \lambda_{max}$.  Then, if a total
number of gravitons $N_\gamma$ is incident over the lifetime of
the experiment, the  criterion for detecting a single particle
becomes:
\begin{equation}
\frac{6\,\sigma }{ \pi}\,\frac{M_d}{ m_p \ell^2}\, N_\gamma \ge 1
\; .
\end{equation}
$N_\gamma$  can be expressed in terms of the source's graviton
luminosity $L_\gamma$ and distance $R$ as
\begin{equation}
N_\gamma = \frac{L_\gamma}{4\pi R^2}\frac{A_d
\tau_s}{\epsilon_\gamma}.
\end{equation}
Here, $A_d = \pi \ell^2/6$ is the effective cross-sectional area
of the detector (i.e., the cross section averaged over possible
paths), $\tau_s$ is the time the source is on and
$\epsilon_\gamma$ is the graviton energy. Thus,
\begin{equation}
\frac{\,\sigma }{4\pi}\,\frac{M_d}{ m_p}\, \frac{L_\gamma}{R^2}
\frac{\tau_s}{\epsilon_\gamma} \ge 1 \; .\label{dctau}
\end{equation}
Reasonably, $\tau_s \le M_s/L$ by energy conservation, where $M_s$
is the mass of the source and $L$ its total luminosity. We also
expect that any astrophysical source emits only a small fraction
of its energy as gravitons.  Letting  $f_\gamma \equiv L_\gamma/L$
yields
\begin{equation}
\frac{f_\gamma \,\sigma }{4\pi}\,\frac{M_d}{ m_p}\,
\frac{M_s}{R^2} \frac{1}{\epsilon_\gamma} \ge 1 \; .
\end{equation}
Thus far this is almost completely general. We now assume that the
detector should not be so large as to undergo gravitational
collapse to greater than atomic densities. It is well
known\cite{CR79} that the requirement that gravitational forces do
not overwhelm the electrostatic forces supporting the object gives
a maximum detector mass of about the mass of Jupiter: $M_d \sim
(\alpha/\alpha_g)^{3/2}m_p$, where $\alpha = e^2/\hbar$ and
$\alpha_g \equiv Gm_p^2/\hbar$ are the fine structure and
gravitational fine structure constants, respectively.  And so,
finally, the criterion for detecting a graviton becomes:
\begin{equation}
\frac{ f_\gamma\,\sigma }{4\pi}\,\left(\frac{\alpha }{\alpha_g
}\right)^{3/2}\, \frac{M_s}{R^2} \frac{1}{\epsilon_\gamma} \ge
 1 \; .\label{dc}
\end{equation}
The crux of the problem is to determine $\sigma$ for a given
process and $f_\gamma$.
%-----------------------------------------------------------------------------
\section{Cross sections: Heuristic Arguments}
 \setcounter{equation}{0} \label{heuristic}
%-----------------------------------------------------------------------------
Whenever dealing with quantum gravity, it is fruitful to enlist
analogies with ordinary quantum mechanics and electromagnetism,
and we shall do so throughout. In choosing an experiment, and
hence a cross section to calculate, we also follow this stratagem
and attempt to learn from history.  A century ago two experiments
were crucial in establishing the particle nature of light: the
photoelectric effect and Compton scattering of photons by
electrons.  The photoelectric effect of course presented severe
difficulties for classical electromagnetic theory, which predicted
that the energy of the photoelectrons should increase with the
intensity of the incident light, but should be independent of the
light frequency, neither of which proved true.  Einstein explained
everything by introducing light quanta and postulating the
celebrated one-liner: $E = h\nu - W$, which gave the
photoelectrons' energy in terms of the frequency of the quantum
and the work function, the energy needed for the electron to
escape the surface of the detector. The number of photoelectrons
simply became equal to the number of incident quanta.  Whether the
formula actually held, though, was debated for ten years as
physicists plotted $E$ versus $\nu$ to get the slope $h$, but in
1916 Millikan was able to announce that ``Planck's $h$ has been
photoelectrically determined with a precision of about 0.5\% to
have the value $h = 6.57 \times 10^{-27}$"\cite{Pais82}.

With perfect hindsight, a devil's advocate would point out that
the photoelectric effect is not so conclusive.  In  semi-classical
derivations of the photoelectric cross section (\S\ref{matrix}),
such as Schiff's\cite{Schiff49}, the electromagnetic field is
never quantized. Thus, although the computed cross section gives
good agreement with experiment, nothing in the mathematics implies
a photon. There are two answers to this rather subtle point: one
is that the photoelectric effect is a fact independent of the
derivation of the cross section, and it still can't be explained
by classical physics.  The other is that we will accept the
existence of a quantum of light (or gravity) if the detector can
``fire" when there are fewer than one quantum in the detector at a
given instant. We make this statement more precise in \S
\ref{discussion}, after we have estimated expected fluxes, but
will consider a ``gravito-electric" experiment in which gravitons
knock electrons from a detector to be a viable candidate.

Compton scattering might be considered a more conclusive
demonstration of the existence of photons, because the change in
wavelength under scattering is calculated by treating the X-rays
as nothing more than billiard balls with quantum-mechanical energy
$E = h\nu$ and momentum $p = h/\lambda$. Thus, we can also imagine
a ``gravito-Compton" experiment in which a graviton is scattered
off an electron and one attempts to detect the electron's recoil
energy. (Detecting the scattered graviton itself, in analogy to
the usual detection of Compton-scattered photons, is not kosher
because it only returns the original problem of detecting the
graviton.)

The gravito-Compton effect presents its own subtleties. The
ordinary Compton scattering cross section is $\sigma_C \sim
(e^4/m^2)$, where $e$ and $m$ are the electron's charge and mass.
One might naively assume by the dimensional substitution $e
\rightarrow {\sqrt G} m$ that gravito-Compton scattering should
have cross section $\sigma \sim G^2m^2$. Indeed, numerous authors
in the 1960s and 1970s considered scattering of gravitons off
neutral scalar particles. Dewitt, for example\cite{DeWitt67},
gives for the nonrelativistic limit the differential scattering
cross section
\begin{equation}
\frac{d\sigma}{d\Omega}= \frac{G^2m^2}{sin^4 (\frac{\theta}{2})}
\left[cos^8 (\frac{\theta}{2}) + sin^8 (\frac{\theta}{2})\right]\,
, \label{gc}
\end{equation}
with an analogous relativistic limit. Perhaps to one's surprise,
although the $G^2m^2$ dutifully appears, (\ref{gc}) resembles the
Rutherford scattering cross section with its $\theta = 0$
divergence rather than the Thomson cross section
$\frac{8\pi}{3}(e^4/m^2)$, which is the nonrelativistic limit of
the Compton cross section. This is a hint to beware analogies
between electromagnetism and gravity.  To the best of our
knowledge the scattering of gravitons off true fermions has not
been calculated, but conversion of gravitons to photons by
scattering off charged scalar particles has\cite{Logi77},with the
result that for unpolarized incident radiation:
\begin{equation}
\frac{d\sigma}{d\Omega}=
\frac{e^2G}{4\pi}cot^2\left(\frac{\theta}{2}\right) \left[sin^2
2\phi + cos^2\theta cos^2 2\phi\right].\label{gpc}
\end{equation}
This expression also diverges, but is larger than (\ref{gc}) by a
factor of $\alpha/\alpha_g (m_p/m)^2 \approx 4 \times 10^{42}$. As
to why these cross sections diverge, the usual statement is that
it is due to the long-range nature of the gravitational and
electromagnetic force, but the different forms of the divergence
or lack thereof in ordinary Compton scattering show that more
detailed considerations play a role.

Because the total cross sections of (\ref{gc}) and (\ref{gpc}) are
formally infinite, in practice one must put in a cutoff in the
$\theta$ integration. This can be accomplished by relating
$\theta$ to the initial  graviton and final electron energies
 in the usual Compton scattering formula; however certainly
for (\ref{gc}) any such detail is irrelevant because $G^2m^2 \sim
10^{-110} cm^2$.  In other words, $G^2m^2 \sim
10^{-45}\ell_{pl}^2$, where $\ell_{pl}\sim 10^{-33} cm$ is the
Planck length. Regardless of what other numbers one inserts into
(\ref{dc}), the detection criterion cannot be satisfied. Gravitons
cannot be detected by Compton scattering of gravitons off neutral
particles.

All hope is not lost, however.  One might reasonably assume the
enormously larger (\ref{gpc}) must be the nonrelativistic result
for true electrons and contemplate using it as the basis for an
experiment. In fact no speculation is required.  For any
``gravito-atomic process," the actual cross section turns out to
be, within numerical factors, the square of the Planck length, in
our units $G\hbar \sim 10^{-66} cm^2$, comparable in size to
(\ref{gpc}) but without divergences. To get some idea of how this
comes about, consider an idealized gravitational wave detector,
consisting of two balls on a spring that behave as damped harmonic
oscillators under the passage of a gravitational wave:
\begin{equation}
\ddot\xi + \dot\xi/\tau_o + \omega_o^2\xi = acceleration,
\label{accel}
\end{equation}
where $\xi$ is the displacement of the balls.  The basic
definition of differential cross section is:
\begin{equation}
d\sigma = \frac{P(ergs/s)}{I(ergs/s/cm^2)}\,. \label{diffcs}
\end{equation}
By computing via (\ref{accel}) the power absorbed in the detector,
$P$, and dividing it by the incoming flux $I \sim h^2\omega^2/G$
of the gravitational wave (see \S\ref{ioncs}), is not difficult to
show\cite{Misner73} that near resonance ($|\omega \pm \omega_o| <<
\omega_o$) the cross section is given by the Lorentzian
\begin{equation}
 \sigma = \frac{\pi G
ML^2(\omega_o^2/\tau_o)sin^4\theta}{(|\omega| - \omega_o)^2+
(1/2\tau_o)^2} \;.
\end{equation}
The mean cross section is just $\frac{1}{\omega_o} \int \sigma
d\omega$, which gives $\sigma_{avg} \sim G ML^2 \omega_o$. If we
now assume that $L = a = \hbar^2/me^2$, the Bohr radius, $M =m$,
the electron mass, and impose the Nicholson-Bohr quantization
condition $m\omega a^2 = \hbar$, then we immediately have
$\sigma_{avg} \sim G\hbar$,  the square of the Planck length.
Notice that the ``atomic process" enters only in determining the
size of $a$; if the atom were bound by gravity, the Nicholson-Bohr
condition would still apply, but  but $a$ would be about $10^{39}$
times larger and $\omega$ about $10^{78}$ times smaller.

Consequently, the ``Planck-length squared" cross section stems
directly from the imposition of angular momentum quantization onto
the detector and nothing more.  This result has the immediate
consequence that it is physically impossible to detect a single
given graviton with high probability.  Detection criterion
(\ref{sdc}) can now be written approximately as
\[
\frac{M}{m_p\ell^3}\; \ell_{pl}^2 \; \ell = \frac{M}{\ell}\;
\frac{\ell_{pl}^2}{m_p\ell} \ge 1
\]
for detector mass $M$ and size $\ell$.  However, $M/\ell < 1$ for
any object that is not a black hole.  If we take $\ell_{pl} = 1$,
then $m_p = 10^{-19}$. The smallest conceivable detector is
presumably about the size of a proton, or $\ell \sim 10^{20}$,
making the inequality impossible to satisfy.  Smolin \cite{Sm85}
has argued that this conclusion holds for any physical detector
and that therefore it is impossible for gravitational radiation to
be in thermodynamic equilibrium with it surroundings.

However, this still leaves open the possibility that given a large
flux of gravitons, a suitable detector might be able to detect one
or more of them. Therefore, heartened by a non-divergent cross
section that is forty some orders of magnitude larger than it
could have been, we proceed to more detailed considerations.

%-----------------------------------------------------------------------------
\section{Matrix Elements}
 \setcounter{equation}{0} \label{matrix}
%-----------------------------------------------------------------------------
The near universal prescription for the  quantum mechanical
calculation of cross sections is to select an interaction
Hamiltonian between the particle and field, then employ
perturbation theory to calculate the transition probability
$P_{ab}$ between individual states:
\begin{eqnarray}
P_{ab}&\propto &\frac{2\pi}{\hbar} |<b| H |a>|^2\nonumber \\
 & \equiv  & \frac{2\pi}{\hbar}|\int\Psi^*_b H\Psi_a d^3x|^2, \label{Gammaqm}
\end{eqnarray}
where $\Psi_a$ and $\Psi_b$ are  the  initial and final
wavefunctions. Multiplying by the density of final states $\rho =
dn/dE$ gives the famous golden rule and the total transition rate
\begin{equation}
\Gamma = \frac{2\pi}{\hbar}|<a|H|b>|^2\rho \, .\label{gr}
\end{equation}
The differential cross section $d\sigma$ for the process under
consideration then follows immediately from the fundamental
definition (\ref{diffcs}) as the ratio between $\hbar\omega\Gamma$
and the incoming flux, $I$. We will follow this prescription, but
for gravity it turns out that the determination of  the
Hamiltonian and consequently
 $\Gamma$ does involve subtleties, and  we will
 try to clarify them.  Doing so  not
only proves necessary to get the correct $\sigma$ but  presents a
few surprises.

Consider first the calculation of the matrix element for the
electromagnetic field via the standard semi-classical
prescription.  In this approach one begins by writing down  the
classical interaction Hamiltonian $H = -e/m\,\bf A\cdot p$, where
$\bf A$ is the  vector potential and $\bf p$ is the electron
momentum. To compute $|<b|H|a>|$ textbooks discuss replacing $\bf
p$ by the momentum operator $-i\hbar\nabla$, but in fact what is
usually done is to apply the correspondence principle in reverse:
$-i\hbar\nabla \rightarrow {\bf p} \rightarrow md{\bf r}/dt$, by
which point both $\hbar$ and quantization of the field have
dropped out of the problem. Additionally,  ${\bf r}$ in the
``energy representation" is taken to have time dependence
$e^{i\omega_{ab} t}$, where $\hbar\omega_{ab}$ is the energy
difference between two levels $a$ and $b$, implying that ${\bf p}
= im\omega_{ab}{\bf r}$. One moreover assumes that the field
itself is varying sinusoidally, i.e., ${\bf A = A_o} e^{i(\omega t
- {\bf k\cdot r})}$.   The time-independent part of the matrix
element thus becomes
 $ A_o e\omega_{ab} |<b|e^{i\bf k\cdot r}{\bf r}|a>|$. Usually, the wavelength in the problem
 is much larger than atomic dimensions, allowing one to make the ``dipole
 approximation," $e^{i\bf k\cdot r} = 1$.  And so, the matrix element one
actually computes is simply $|<b|H|a>| = A_o e\omega_{ab} |<b|{\bf
r}|a>|$.\\

For the gravitational case, we  copy the procedure, considering
the interaction of an electron with a gravitational wave.  A plane
gravitational wave can be viewed as a weak perturbation
$h_{\mu\nu}$ traveling on a flat  background $\eta_{\mu\nu}$, such
that the full spacetime metric becomes:
 \begin{eqnarray}
g_{\mu\nu} = \eta_{\mu\nu} + h_{\mu\nu}.\nonumber
 \end{eqnarray}
 In analogy to the electromagnetic case,
we take for a monochromatic gravitational plane wave $h_{\mu\nu} =
he_{\mu\nu}e^{i(\omega t - \bf k \cdot r)}$, where $e_{\mu\nu}$ is
the polarization tensor.  Because the perturbations are assumed
weak, one takes the amplitude
 $h << 1$.

To illustrate how to get an interaction Hamiltonian, it is perhaps
easiest to proceed as follows: Recall that in special relativity,
the Minkowski metric can be written $d\tau^2 =
-\eta_{\mu\nu}dx^\mu dx^\nu = dt^2(1-v^2)$.  This leads to the
free-particle Lagrangian $L = -m\sqrt{ 1-v^2}$. For an interaction
with a gravitational wave we replace $\eta_{\mu\nu}$ by
$g_{\mu\nu} = \eta_{\mu\nu} + h_{\mu\nu}$ to get
\begin{equation}
L = -m{\sqrt{ 1-v_iv^i - h_{\mu\nu}v^\mu v^\nu}},
\label{lagrangian}
\end{equation}
Here, $v^\mu \equiv dx^\mu/dt$; we use the summation convention
throughout and Latin indices refer to spatial components. Although
this Lagrangian is not manifestly covariant, it serves for our
purposes.

Taking $\partial L/\partial v^\alpha$ yields by definition the
canonical momentum $\pi_\alpha$.  In the nonrelativistic limit $v
<< 1$.  To lowest order in $v$ the canonical momentum becomes:
\begin{eqnarray}
\pi_\alpha = p_i\delta^i_\alpha + h_{\alpha\beta}p^\beta,\nonumber
\end{eqnarray}
where $\delta^i_\alpha$ is the Kronecker delta and $p_\alpha =
mv_\alpha$ is the ordinary momentum. By definition, the
Hamiltonian is $H = \pi_\alpha v^\alpha - L$.  Working again to
lowest order gives:
\begin{eqnarray}
 H = \frac{p_i p^i}{2m} +
\frac{1}{2}\frac{h_{\alpha\beta} p^\alpha p^\beta}{m}. \nonumber
\end{eqnarray}
(Here we have ignored a physically irrelevant constant $m$.) If we
define $T^{\mu\nu} \equiv  m v^\mu v^\nu$ to be the
 classical stress-energy tensor for an electron (integrated over volume)
the interaction Hamiltonian is then
\begin{equation}
 H_{int} = \frac{1}{2}h_{\mu\nu}T^{\mu\nu},
\label{Hamiltonian}
\end{equation}
and we see that $T^{\mu\nu}$ plays the same role as ${\bf p}$ did
in the in the electromagnetic case,  while $h_{\mu\nu}$
corresponds to ${\bf A}$.  This derivation agrees with the more
general result found in \cite{Dyson69,Weinberg72}.

Now, the polarization tensor $e_{\mu\nu}$ has only two independent
components.  For the moment let us  make the standard choice
$e_{11} =  -e_{22}$ and $e_{12} =  e_{21}$. If we again make the
dipole approximation, then sandwiching $H_{int}$  between the
initial and final state functions gives the required matrix
element $1/2<b|e^{ij}p_ip_j|a>$. To evaluate this expression
exactly, one should let $p\rightarrow -ih
\partial/\partial x$. However, if following the semi-classical
procedure, we set $p = m\dot x$, then we have $m^2<b|e^{ij}\dot
x_i \dot x_j|a>$, which dimensionally is $m^2\omega^2<b|e^{ij}x_i
x_j|a>$. Indeed, for the harmonic oscillator wavefunction, the
expectation value $m^2\omega^2<a|x^2|a>$ is identically equal to
$<a|p^2|a>$, and one can also show that the matrix element
$m^2<b|\dot x_i \dot x_j|a> = m^2\omega^2<b|x_i x_j|a>$ for any
case in which all the transition frequencies $\omega$ are the
same. Thus we expect:
\begin{equation}
\begin{array}{lll}
<b|H|a> &\approx& \frac{mh\omega^2}{2}\int\Psi^*_b e^{ij}x_i x_j\Psi_a d^3x \vspace{2mm}\\
&=& \frac{mh\omega^2}{2}e^{ij}D_{ij},\label{me}
\end{array}
\end{equation}
where
\begin{equation}
D_{ij} \equiv \int\Psi_b x_i x_j\Psi_a d^3x. \label{Dqm}
\end{equation}

 In principle, all we need to do now to get $\Gamma$ is insert the
relevant wavefunctions into $D_{ij}$ and calculate $|<b|H|a>|^2$.
The matter is not so simple, however. First, for any absorption or
scattering process the graviton can be incident from any
direction. Therefore, we need to square (\ref{me}) for each
polarization, average the results over the unit sphere and sum
them. To do this, take $|e_{ij}| = 1/\sqrt2$. Then,
\begin{eqnarray}
<| D_{ij} |^2> \equiv  \frac{1}{4\pi}\int d\Omega \left[2|\Psi_b
xy\Psi_a|^2 + \frac{1}{2}|\Psi_b (x^2 - y^2)
\Psi_a|^2\right].\nonumber \label{avg}
\end{eqnarray}
In BR, we give a quick way to evaluate this expression.  The
result is:
\begin{equation}
<|D_{ij}|^2>  = \frac{2}{5}(D_{ij}D_{ij} -
\frac{1}{3}|TrD|^2).\label{finalavg}
\end{equation}
In what follows, we will thus assume that
 \begin{equation}
|<b|H|a>|^2 = \frac{1}{4} \frac{m^2h^2\omega^4}{4}<|D_{ij}|^2>,
\label{me2}
\end{equation}
where $<|D_{ij}|^2>$ is given by Eq. (\ref{finalavg}).  The
 extra $1/4$ in this equation gives an exact expression in a sense
 explained in the Appendix. It
arises, in part, when demanding that detailed balance be satisfied
among the absorption, spontaneous and stimulated emission rates.
The appearance of this factor also helps resolve some
discrepancies in the literature and leads to some fundamental
issues regarding gauge transformations in general relativity.
Because these considerations are somewhat more technical than the
general discussion and do not seriously affect the
 numerical result, we leave greater detail to the Appendix
and to BR.  Here, we proceed to  calculate a cross section. \\

%-----------------------------------------------------------------------------
\section{Ionization cross section}
 \setcounter{equation}{0} \label{ioncs}
%-----------------------------------------------------------------------------

Despite  our devil's advocate insistence that nothing in the
previous discussion implied the existence of quanta, we again
defer comment until \S\ref{discussion} and
 choose to
 calculate the gravito-electric (ionization) cross section  of atomic hydrogen.
 Although some metals have work functions that are somewhat smaller than hydrogen's, hydrogen's is
 typical and the element does compose most of the universe; for this reason we chose a
hydrogen detector in \S\ref{detection}.  It turns out that in the
energy regime of greatest interest, the result is not sensitive to
the work function.

 As in the standard calculation of the photoelectric cross
 section\cite{Schiff49}, we take the initial state to be the ground (1s) state  of
hydrogen and for the final state a plane wave with momentum $\bf
k$:
\begin{eqnarray}
\Psi_a =\frac{1}{\sqrt{\pi} a^{3/2}}e^{-r/a} &;&\Psi_b =
\frac{1}{L^{3/2}}e^{i{\bf k \cdot r}},\label{ionwavfcns}
\end{eqnarray}
where $L$ is the normalization constant.  The use of the plane
wave final state (the Born approximation) is reasonably good for
large final momentum, or $a^2k^2 > > 1$, which is again the case
of most interest.

Before one computes the matrix element with these wavefunctions,
however, there is a complication.  As in the ordinary
photoelectric effect (see Schiff, p. 287), one should average the
direction of the incident graviton over all angles.  This average,
though, is exactly the one we have already performed, leading to
Eq. (\ref{finalavg}), and therefore, in evaluating the matrix
elements expression (\ref{me2}) should be  used for\\
$|<b|H|a>|^2$.

 This actually
simplifies matters greatly, because with the gravitational field
averaged over all directions, we may assume that the electron is
ejected along the z-axis and that $\theta$, the angle between
${\bf k}$ and ${\bf r}$,  is the usual polar angle. Once again the
integrals can all be carried out analytically in terms of
elementary functions and the result is:
\begin{equation}
|<b|H|a>|^2 = \frac{3(32)^2\pi}{5} \frac{h^2\omega^4 m^2a^7
(a^4k^4)}{L^3 (1+a^2k^2)^8}
\end{equation}

The total absorption rate $\Gamma$ follows directly from the
golden rule (\ref{gr}). With the standard expression for the
density of states $\rho= L^3/(2\pi\hbar)^3 km d\Omega$, this gives
\begin{equation}
\Gamma = \frac{3(32)^2}{4 \cdot 5\pi} \frac{h^2\omega^4
m^3a^{11}k^5} {\hbar^3(1+a^2k^2)^8} d\Omega
\end{equation}
The differential cross section now follows from (\ref{diffcs}) as
$d\sigma = \hbar\omega\Gamma/I$. For
 a gravitational plane wave, the equivalent of the Poynting flux is\cite{Misner73}
 $I = \omega^2 h^2 e_{ij}e^{ij}/16\pi G = \omega^2 h^2/8\pi G $, yielding
\begin{equation}
d\sigma = \frac{6
(32)^2}{5}\frac{G\omega^3m^3a^{11}k^5}{\hbar^2(1+a^2k^2)^8}
d\Omega
\end{equation}
But the photoelectric equation requires $E = (\hbar k)^2/2m =
\hbar\omega - e^2/2a$, where $e^2/2a$ is the hydrogen work
function. Hence, $1+k^2a^2 = 2m\omega a^2/\hbar$. Inserting this
into the above and integrating yields, finally, the total cross
section
\begin{equation}
\sigma = \frac{3072\,\pi}{5}\frac{\hbar G (ak)^5}{(1+a^2k^2)^5}
\,.\label{sigma}
\end{equation}
As promised, $\sigma \sim \hbar G$ to ``within factors of order
unity," but is strongly dependent on $ak$. (Indeed, for $ak >> 1$,
the dependence is the same as in the ordinary photoelectric
effect.)  We attempt to evaluate this factor in the following
section.

%-----------------------------------------------------------------------------
\section{Graviton Production}
 \setcounter{equation}{0} \label{production}
%-----------------------------------------------------------------------------
The second half of the task is to determine the graviton
luminosity of astrophysical sources, that is  estimate $f_\gamma$
in Eq.(\ref{dc}).

There are only a few conceivable sources of gravitons: spontaneous
emission of gravitons from neutral hydrogen, black hole decay,
bremsstrahlung from electron-electron collisions in stellar
interiors and conversion of photons to gravitons by interstellar
magnetic fields. We examine each in turn.

Spontaneous emission of gravitons, such as that from the $3d
\rightarrow 1s$ state of hydrogen discussed in the Appendix, will
not produce gravitons energetic enough to ionize hydrogen.  One
can, alternatively, imagine a gravito-Compton experiment with the
cross section $\sigma \sim e^2G$ from \S\ref{heuristic}. We leave
it to the reader to show that, with the calculated decay rate
$\Gamma \approx 5 \times 10^{-40} s^{-1}$, even if one assumes
that the entire mass of the galaxy, $M_g \approx 10^{50} M_{pl}$
resides in hydrogen in the $3d$ state at an average distance of 30
kiloparsecs, $R \approx 10^{56} \ell_{pl}$, the detection
criterion  falls orders of magnitude short of being satisfied.
Detection of gravitons from the spontaneous emission of hydrogen
 is impossible.\\

Graviton production by black hole decay is in some sense more
promising. The Hawking temperature of black holes is $T_{bh} \sim
\hbar/kGM$. To evaporate $10 eV$ or higher energy gravitons
requires black holes of $M \le 10^{27} M_{pl}$ or  $M \le 10^{22}
g$. As this is far less than a stellar mass, such black holes
would necessarily be primordial. Although theoretical prejudice
may be
 aligned against primordial black holes, observational
constraints on this mass range of PBHs are almost
nonexistent\cite{Carr04} and in principle they could make up most
of the universe's dark matter. On the other hand, PBHs with $M \le
10^{15} g$, which would emit particles with energy $\ge 10^8 eV$,
 would have evaporated by the current age of the universe.
Constraints on such PBHs are much tighter due to the distortion
they produce on the X-ray background\cite{Carr04}. Let us then
assume PBHs with $10^{15} g < M < 10^{22} g$  make up the entire
 mass of the galaxy.

Detection criterion (\ref{dc})  requires to an order of magnitude
\begin{equation}
10^{53}  f_\gamma\,\sigma \frac{M_s}{R^2}\frac{1}{\epsilon_\gamma}
\ge
 1 . \label{dcnum}
\end{equation}
To determine $f_\gamma$ over the range of interest, note that the
temperature of a $10^{22} g$ black hole is too low to evaporate
anything but massless particles: neutrinos, photons and gravitons.
By equipartition one expects approximately equal numbers of these
species to be radiated, but the higher spin of the graviton
suppresses its emission compared to the neutrino and photon.
Page's detailed calculations\cite{Page76} have shown in this case
that $f_\gamma = .02$. For PBHs with temperature above the
electron emission threshold, $f_\gamma$ drops only to $.01$, so it
can be considered a constant $\approx .01$ over the entire mass
range.

To estimate $M\sigma/\epsilon_\gamma$, first note that from the
photoelectric equation we have $a^2k^2 = (\omega/\omega_{Ry}) -
1$, where the Rydberg frequency $\omega_{Ry} \equiv
\hbar/(2ma^2)$. Thus $ak \approx (\epsilon_\gamma)^{1/2}$ when the
graviton energy $\epsilon_\gamma$ is in Rydbergs. The mass
spectrum is expected to be dominated by the smaller
holes\cite{Carr04} of $M\sim 10^{15} g$, which emit gravitons at
about $10^7 Ry$ or $ak \sim 10^{3.5}$. If we assume all the holes
are in this mass range, the $(ak)^5$ in the denominator of the
ionization cross section reduces it to only $\sigma \sim 10^{-14}
\hbar G$. The high graviton energy $\epsilon_\gamma \sim 10^7 Ry
\sim 10^{-20} \epsilon_{pl}$ is similarly unfavorable in criterion
(\ref{dcnum}). If we again take $M_g \approx 10^{50} M_{pl}$ and
$R \approx 10^{56} \ell_{pl}$, we find that the number of
detections over the source lifetime for a Jupiter mass detector is
\begin{equation}
N_d \sim 10^{-5}.
\end{equation}
Properly one should  average $(M\sigma/\epsilon_\gamma)$ over the
mass spectrum of PBHs, or equivalently the lifetime of the source
$\tau_s$ in Eq. (\ref{dctau}), but  the estimate we've just done
should be adequate. Larger holes give a larger cross section, but
their lifetime $\tau \sim M^3$ is so much enormously larger than
the age of the universe, that the detections per age of universe
is negligible. We should also mention that our ionization cross
section is not valid in the relativistic regime, but a proper
calculation may only make things worse; if so this result does
indicate that even if primordial black holes exist, they do not
produce a sufficient quantity of gravitons
to be detected. \\

A more definite astrophysical source of gravitons is gravitational
bremsstrahlung produced by electron-ion or electron-electron
collisions in stellar interiors. Gravitational bremsstrahlung has
been studied by numerous authors (see Weinberg\cite{Weinberg72}
and Gould\cite{Gould85}).  It is calculated by much the same
procedure as in electrodynamics, however there the radiation is
primarily dipolar, whereas as discussed in the Appendix,
gravitational radiation is quadrupolar.  In this case
 one  calculates radiated gravitational energy in analogy to
 electromagnetic  quadrupole radiation, by taking the
the square of the third time derivative of the moment of inertia
tensor, $D_{ij}(t) = \mu x_i(t)x_j(t)$, for reduced mass $\mu$. In
the zero-frequency limit, one finds that the energy radiated per
unit frequency is:
\begin{equation}
\frac{dE}{d\omega} = \frac{8G}{5\pi}\mu^2 v^4
sin^2\theta,\label{dEdOm}
\end{equation}
where   $\theta$ is the scattering angle of the electron.

If two scatterers with charges $e_a$ and $e_b$ undergo Coulomb
scattering with impact parameter $b$, the deflection angle will be
$\theta \approx e_ae_b/(b\mu v^2)$.  The maximum frequency of
emitted radiation will be $\omega_{max} \sim v/b \sim \mu
v^3\theta/(e_a e_b)$. The important point is that the frequency of
emitted radiation depends on $\theta$. Integrating the above to
$\omega_{max}$ yields for small angles
\begin{equation}
E \approx \frac{8G}{5\pi}\frac{\mu^3 v^7 sin^3\theta}{e_a e_b}.
\end{equation}
The   gravitational power radiated per unit volume is found simply
by multiplying this expression by the number of  collisions per
unit time and volume, or
\begin{eqnarray}
P = v \sum_{a,b} n_a n_b \int E  \frac{d\sigma_{r}}{d\Omega}
d\Omega,\nonumber
\end{eqnarray}
where $n_e$ and $n_b$ are the number density of scattering
species. In the nonrelativistic regime $d\sigma_r/d\Omega$ can be
taken to be the differential Rutherford scattering cross section
\begin{eqnarray}
\frac{d\sigma}{d\Omega} = \frac{e_a^2e_b^2}{4 v^4 \mu^2
sin^4{\theta/2}}.\nonumber
\end{eqnarray}
For a Maxwellian velocity distribution the power per unit volume
works out to be
\begin{equation}
P \approx \frac{64 \pi G}{5} \frac{15 (kT)^2}{\mu} \sum_{a,b}n_a
n_b e_a e_b,\label{power}
\end{equation}
Multiplying P by the volume of the star gives the power radiated
in gravitational bremsstrahlung.  We note that this formula
differs somewhat from Weinberg's Eq. (10.4.31), which contains a
lower-limit $\theta$ cutoff, $\ln \Lambda$, due to the divergence
in the Rutherford cross section. His expression, however, neglects
the fact that the upper limit of integration for $\omega$ depends
on $\theta$ (and vice versa). When carried out as we have, the
$\ln \Lambda$ does not appear.

In any case, we present this derivation primarily  for
illustrative purposes because  Eq. (\ref{dEdOm}) is  only valid
for low-energy electrons. Stellar interiors, where $T > 10^7 K$
($kT > KeV$), require the opposite extreme; and it is in this
limit where the Born approximation used in the calculation of the
ionization cross section is valid. Because Gould has treated in
some detail the astrophysical regime, we here merely quote  his
results. He finds that for the Sun, $L_{\gamma *} \sim 7.9 \times
10^{14}\, ergs\, s^{-1}$ implying that $f_{\gamma *} \sim 2 \times
10^{-19}$. For white dwarfs he finds $L_{\gamma wd} \sim 10^4
L_{\gamma *}$, but for a typical white dwarf, $L_{wd} \sim
10^{-2}L_*$, implying that $f_{\gamma wd} \sim 10^{-13}$. For
neutron stars he gets $L_{\gamma ns} \sim 10^{25} \, ergs\,
s^{-1}$.  The total thermal luminosity of neutrons stars is less
well determined, because much of the luminosity is due to
rotational energy.  Recent observations\cite{Kaminker05}, however,
give a thermal luminosity $L_{ns} \sim 10^{31-34}ergs\, s^{-1}$,
implying $f_{\gamma ns} \sim 10^{-8}$, very roughly.

The number of graviton detections from bremsstrahlung is now
easily estimated. If $T \sim 1 KeV$, then $ak \sim 10$, which
gives for the ionization cross section $\sigma \sim .01 \hbar G$.
With $M$ and $R$ as before, and making the assumption that all the
mass of the galaxy resides in Main Sequence stars, Eq.
(\ref{dcnum}) yields
\begin{equation}
N_\gamma \sim 10^{-5},
\end{equation}
 ruling out detection of gravitons, but if a
substantial fraction of the galaxy resides in white dwarfs or hot
neutron stars (for the latter: $T \sim 10^9; ak \sim 100$),
\begin{equation}
N_{\gamma wd} \sim 10; \ N_{\gamma ns} \sim .1 , \label{Ngal}
\end{equation}
giving a faint hope.  Unfortunately, the uncertainties in the
various quantities preclude more precise answers.  In this regard
we point out that in this calculation and  the previous, the
numbers are not significantly changed by going to the mass and
radius of the observable universe, due to the inverse square law;
this is a manifestation of Olber's paradox.  However, one can do
better by merely parking the detector at 1 AU from the source,
which yields
\begin{equation}
N_{\gamma *} \sim 10^{3};\ \ N_{\gamma wd} \sim 10^{9} ;\ \
N_{\gamma ns} \sim 10^{7}. \label{NAU}
\end{equation}\\
At the tidal disruption radius $r_T \approx (M_*/M_J)^{1/3}R_J
\approx 30 R_J$, these numbers can be pushed to
\begin{equation}
N_{\gamma *} \sim 10^{7};\ \ N_{\gamma wd} \sim 10^{13} ;\ \
N_{\gamma ns} \sim 10^{11}
\end{equation}
For a neutron star thermal lifetime of roughly $10^5$
years\cite{Kaminker05},  the last figure gives potentially
$10^{6}$ graviton detections per year.

All these numbers are, however, somewhat optimistic. First of all,
current estimates of the number of neutron stars in the galaxy is
$\sim 10^9$, for a mass fraction $\sim 10^{-3}$.  Furthermore,
detection criterion (\ref{dcnum}) assumed that the upper limit for
the lifetime of a source was $ \tau = M/L$. In real life, for a
main sequence star $\tau \sim 10^{-3}M/L$; for white dwarfs $\tau
< 10^{-5}M/L$; and for neutron stars $\tau_{th} < 10^{-8}M/L$.
Thus, the white dwarf result in (\ref{Ngal}), as well as the MS
and NS results of Eq. (\ref{NAU}) are also ruled out absolutely.
At this stage, the only possibility  for detecting gravitational
bremsstrahlung appears to be putting the Jupiter-mass detector in
close orbit around a white dwarf or neutron star; the latter might
result in as many as
$10^{-2}$ detections per year.\\

A surprising, but perhaps significant, potential source for
gravitons is photon-graviton oscillations due to passage of
photons through the galactic magnetic field.  This mechanism,
analogous to neutrino oscillations, was first discussed by \\
Gertsenshtein\cite{Gert62}.  If $T_{\mu\nu}$ in interaction
Hamiltonian (\ref{Hamiltonian}) contains stress-energy due to a
background magnetic field $\bf B$, as well as to the  magnetic
field $\bf b$ of the photon, then the linearized wave equation for
the gravitational wave $h_{\mu\nu}$ contains both these fields in
the source term. Moreover, $h_{\mu\nu}$ must also satisfy
Maxwell's equations in a gravitational field, which couples the
photon magnetic field to the gravitational wave.  Simultaneous
solution of these equations leads to a wave-packet that oscillates
between the photon and graviton states with mixing length
\begin{equation}
L=(2/G^{1/2}B),
\end{equation}
independent of wavelength. (Here $B$ is the component of the
magnetic field perpendicular to the direction of propagation.) For
$B$ on the order of a Gauss, $L$ comes out to be roughly a
megaparsec. Furthermore,  the oscillation period shows that if a
single photon travels a distance $D$ through the field $B$, it
will emerge as a graviton with probability
\begin{equation}
P=\sin^2 (D/L) = \sin^2 (G^{1/2} BD/2).
\end{equation}
Taking the magnetic field of the galaxy to be $B \approx 5\mu G$
with $D$ the distance to the galactic center, yields $P \sim
10^{-15}$.  This is equivalent to $f_\gamma \sim 10^{-15}$ in the
previous calculations.  For UV photons with $\epsilon_\gamma
\approx 2 \, Ry$ the detection criterion gives
\begin{equation}
N_\gamma \sim 10^5.
\end{equation}
The situation improves if one places the detector 1 AU from a
neutron star with a magnetic field $B \sim 10^{12}$.  We take $D
\sim 10 km$.  Then P $\sim 10^{-14}$ and the detection criterion
gives
\begin{equation}
N_\gamma \sim 10^8,
\end{equation}
assuming $\epsilon_\gamma \sim 1KeV$ (which may be somewhat high
for an estimate of neutron star surface temperature). If one put
the detector at the tidal disruption radius, the number of
detections could be raised to $N_\gamma \sim 10^{12}$. Taking into
account factor of $10^{-8}$ already mentioned, we might expect
$10^{-1}$ detections per year for the Jupiter-mass detector.

In reality, however,  the outlook for the Gertsenshtein mechanism
is less positive.  Large magnetic fields will result in
electron-positron pair production, which produces an effective
index of refraction that lowers the speed of light. This in turn
introduces a coherence length above which the gravitons and
photons will no longer be in phase; the Gertsenshtein process is
quenched.  Dyson has recently shown\cite{Dyson05} that the quench
condition is $DB^2\omega \le 10^{43}$, implying  for neutron stars
that $\omega < 10^{13}$. This is below the ionization threshold
for hydrogen; neither is it
  of great astrophysical interest.\footnote{We have recently learned that
Zel'dovich arrived at a similar conclusion regarding the quenching
of the Gertsenshtein mechanism in 1973. See Ya. B. Zel'dovich,
{\it Sov. Phys. JETP}\ {\bf 38}, 652-653 (1974).} Evidently,
neutron star gravitons produced by the Gertsenshtein mechanism
cannot be detected.

%-----------------------------------------------------------------------------
\section{Discussion}
 \setcounter{equation}{0} \label{discussion}
%-----------------------------------------------------------------------------
We are now in a position to address the issue that we have so far
avoided: given that the calculation of the cross section did not
involve quantizing the field, what is our justification for
claiming that a click in the above detector constitutes detection
of a graviton?  To see the answer, consider the upcoming
generation of gravitational wave detectors, which are expected to
be able to measure a wave amplitude $h \sim 10^{-20}$ at
frequencies of approximately $10^3 Hz$.  From the expression for
the flux $I$ above, this corresponds to roughly $10^{7} eV \,
cm^{-3}$ or about $10^{26} eV$ per cubic wavelength.  At an energy
$\hbar\omega$ per graviton this amounts to approximately $10^{38}$
gravitons per cubic wavelength. In other words, LIGO and its
successors will never be said to have detected anything like a
single graviton. The picture just described is one of a classical
wave, which is what LIGO will have detected, as designed. On the
other hand, the fluxes for the sources we have been considering
amount to $\le 10^{-18} eV$ per cubic wavelength,
 far smaller than the  $10\, eV$ imparted to an
  electron in an ordinary photoelectric
experiment.  Such a situation cannot be reconciled with a
classical picture of a wave, because there is insufficient energy
at a given location to ever eject an electron. Hence, if the
ejected electron is registered, we are entitled to claim we have
detected a single graviton.

This last ``if," however, is a large one.  Throughout we have
assumed an ideal detector, of one hundred-percent efficiency, the
mass of Jupiter.  This is not reasonable.  One must therefore
examine the physics of the detector itself. For an ejected
electron to be registered by a sensor, the sensor should be
located within a mean-free-path  of the ejection point. If the
mean-free-path $\lambda$ for an electron in this case is
determined by the ordinary Compton scattering cross section $\sim
10^{-24} cm^2 $, the density is $\sim 1 g cm^{-3}$ and the
temperature  $T \sim 10 K$, then  $\lambda \sim 1 m$, meaning that
a fraction $\lambda/R_J \sim 10^{-6}$ of all ejected electrons
reach the surface. This essentially rules out any of the above
scenarios that were not ruled out by other factors.

One  does not wish to compress the detector, because the
mean-free-path scales as $\ell^{3}$, but it does suggest building
 detectors with $\ell <
\lambda$.  As one example, imagine a honeycomb
 structure made
of silicon, whose interstices are filled with hydrogen. Both the
silicon and hydrogen would become targets (with presumably
comparable cross sections) and the question becomes, How large can
one make such a detector  before self gravity causes structural
failure?  A straightforward calculation shows that the maximum
radius of such a  detector is give by
\begin{equation}
R = \sqrt{\frac{3\epsilon {v_s}^2}{2\pi\rho G}}
\end{equation}
where $\epsilon$ is the structural strength as a fraction of the
bulk modulus, $v_s$ the velocity of sound, and $\rho$ the density
of the material.  Suppose $ \rho = 0.233\, g \, cm^{-3}$ (i.e.,
0.1 the density of silicon),  $\epsilon \sim 0.03$ and  $v_s = 4
\times 10^6 cm \, s^{-1}$.  This yields  $R \sim 4 \times 10^{8}
\, cm$ with a mass of $M \sim 5 \times 10^{25}g$, about $.01
M_{Earth}$.  With such a device the detection criterion cannot be
met for any of the sources  discussed above and detection of
gravitons is ruled out.

This result, however, does not {\em absolutely}  exclude detection
of gravitons; one can imagine filling the solar system and beyond
with tiny detectors.  At this point, though, the possibilities go
out of sight.

Before that point, we must address two other issues.  The first is
noise.  Any detector needs to be shielded against background
noise.  Two serious noise sources are  neutrinos and cosmic rays.
The cross section for the interaction of neutrinos with matter is
about $10^{-45} cm^2$, or at least twenty orders of magnitude
greater than the gravito-electric cross section.  In a typical
white dwarf,  neutrino emission  exceeds photon
emission\cite{ST83}, meaning that $10^{13-14}$ neutrinos are
emitted for every graviton. Therefore, without shielding, one
would expect $10^{33-34}$ neutrino events for every graviton
event.  A shield should be thicker than the mean-free-path for
neutrinos, which for materials of ordinary density amounts to
light years.  Such a shield would collapse into a black hole.
Unless one can find another way to discriminate against neutrinos,
this appears to make detection of thermal gravitons impossible. In
light of this result, we do not pursue shielding against cosmic
rays, which would activate the detector material, inundating it
with secondary particles.

The second issue we have ignored brings us back  to the
philosophical side. We have assumed that a click in the detector
amounts to a detection of a graviton.  Historically, as mentioned
in \S\ref{heuristic}, Millikan's ``proof" of the existence of
photons rested on measuring the slope of the graph of the
photoelectrons' energy versus frequency, i.e., the determination
of $h$.  If one insists that the existence of gravitons is not
established until $h$ is fished out of the data, then further
obstacles immediately present themselves.  At the very least, one
requires many more detections, in order to plot gravito-electron
energy vs. frequency. Unfortunately, in contrast to Millikan's
situation, we here do not have a monochromatic graviton source. To
deconvolve the signal from the source spectrum  presents
additional hurdles.

In sum, we can say that to detect a single graviton was {\em a
priori} going to be a difficult proposition, but it was not
obvious that it was fundamentally impossible. Although, as we
stated at the outset, we have found no basic principle ruling out
graviton detection, reasonable physics appears to do so. Perhaps
the most interesting aspect of the investigation is that it leads
to some fairly subtle physics, which, as the Appendix shows, has
caused significant confusion in the literature. Certainly, if a
``no graviton" law appears elusive, we do feel entitled to predict
that no one will ever detect one in our universe.\\

{\bf Acknowledgements}. We are grateful to Freeman Dyson, not only
for asking the question that prompted this paper, but for numerous
conversations, enthusiasm and for generously making available
unpublished notes. He has effectively been an invisible third
author, although he should not be held responsible for errors on
our part. We also thank the physics department at Princeton
University where much of this work was carried out.

-----------------------------------------------------------------------------
\section*{Appendix: Spontaneous Emission and Hamiltonians}
\setcounter{equation}{0}
\renewcommand{\theequation}{A.\arabic{equation}}
%-----------------------------------------------------------------------------

The extra factor of one-quarter in  the squared matrix element
(\ref{finalavg}) surfaces during the resolution of a matter
concerning the spontaneous emission of gravitons by hydrogen.
Specifically, Steven Weinberg in his standard text {\it
Gravitation and Cosmology}\cite{Weinberg72} employs semi-classical
methods to compute the spontaneous decay rate from the $3d2
\rightarrow 1s$ state of hydrogen by graviton emission and arrives
at one answer, whereas the authors of the well-known {\it Problem
Book in General Relativity and Gravitation}\cite{Lightman75} use a
field theoretic approach for the same problem and arrive at an
answer $2.1 \times 10^5$ times larger.  This is a significant
discrepancy, particularly in that for the electromagnetic case,
the field-theoretic and semi-classical approaches are known to
give identical results.  Although spontaneous emission did not
directly figure in the body of the paper, it does provide an
important check for our calculations because transition rates must
necessarily satisfy detailed balance, i.e. conservation of energy.
Resolving the discrepancy also leads to some fundamental issues
about the Hamiltonian employed in computing the matrix element.

 Consider, then, the spontaneous
emission of a graviton by the decay of a hydrogen atom from the
$3d$ to the $1s$ state (which one might  conceivably contemplate
as a graviton source). In the semi-classical approach to
conventional quantum mechanics\cite{Schiff49}, one does not use a
Hamiltonian to compute the spontaneous emission  rate. (If the
photon or graviton does not exist before emission, it is unclear
what the interaction is.) Indeed, there is no classical analog to
spontaneous photon emission, but in the semi-classical approach
one nevertheless constructs one by imagining a current flowing
from the upper state to the lower. One assumes the current density
is ${\bf J = \rho v = \rho p}/m$. Next one identifies $\rho$ with
the probability density $e \Psi^*\Psi$ taken over the initial and
final states, such that ${\bf J} = e/m\Psi^*_b {\bf p}\Psi_a$. As
in \S\ref{matrix}, one assumes that ${\bf p} = m{\bf v}= im\omega
{\bf r}$. Then ${\bf J} = ie\omega \Psi^*_b {\bf r}\Psi_a$, and
quantization of the field is avoided. In classical
electromagnetism, the total power radiated by a dipole is
$\frac{4k^2}{3}|I_o|^2$, where $I_o$ is the total current, so from
the spontaneous transition we have for the total radiated power:
\begin{eqnarray}
P = \frac{4k^2e^2\omega^2}{3}\; |\int \Psi^*_b x_i \Psi_a
d^3x|^2.\label{powerdip}
\end{eqnarray}
If we now say that each transition emits a quantum with $E =
\hbar\omega$, where $\omega$ is the transition frequency between
the two states, we can reinterpret (\ref{powerdip}) as the
transition rate
 \begin{equation}
 \Gamma = \frac{P}{\hbar\omega}. \label{Gamma}
\end{equation}

There is no reason to believe this derivation, except that when
the stimulated emission and absorption are independently computed,
the imposition of detailed balance leads to the Planck formula
and, miraculously, a proper field theoretic calculation also leads
to the same result.

Thus as in \S\ref{matrix} we again copy the procedure for gravity,
with one important difference.  The power emitted by a simple
electric dipole is $P = (2/3)\ddot d^2$, where $d = ex$ is the
dipole moment. By analogy, one would expect a gravitational dipole
to emit $P \sim (m\ddot {\bf x})^2$.  Due to conservation of
momentum, however, $m{\bf\ddot x = \dot p} \equiv 0$ for an
isolated system and gravitational dipole radiation does not exist.
The lowest
 radiating moment is the quadrupole, and like an electromagnetic
 quadrupole
 the radiated power  goes as $\dot {\ddot Q}^2$.  With the
 substitution $Q \sim ex^2 \rightarrow \sqrt G mx^2$, for an
 $x$ that varies harmonically in time we have $P \sim Gm^2\omega^6x^4$.

More precisely, when the source dimension is much smaller than a
wavelength (the dipole approximation)  one finds\cite{Weinberg72}
that the emitted power is:
\begin{equation}
P = \frac{2G\omega^6m^2}{5}\left[D^*_{ij}D_{ij} -
\frac{1}{3}|TrD|^2\right], \label{powerquad}
\end{equation}
where the moments are exactly those defined by Eq. (\ref{Dqm}) and
2/5 times the bracketed quantity is exactly the average given by
Eq. (\ref{finalavg}). Eq. (\ref{powerquad}) is essentially the
square of the third time derivative of the moment of inertia; the
spatial average results in the  subtraction of $\frac{1}{3}|Tr
D|^2$.  The important point is that all the matrix elements are
now seen to be the same.

The spontaneous emission rate is now found by plugging
(\ref{powerquad}) into (\ref{Gamma}). For the  $3d2 \rightarrow
1s$ transition  the normalized hydrogen wavefunctions are
\begin{eqnarray}
\Psi_{1s} = \frac{1}{\sqrt{\pi} a^{3/2}}e^{-r/a}  &;&
\Psi_{3d2}=\frac{1}{162\sqrt \pi}
\frac{1}{a^{3/2}}\left(\frac{r^2}{a^2}\right)e^{-r/3a}sin^2\theta
e^{2i\phi}\label{wavefcns1}
\end{eqnarray}
and the integrals can all be evaluated analytically in terms of
elementary functions.  Weinberg\cite{Weinberg72} apparently
calculates this transition by such a direct substitution (see his
Eqs. 10.8.5-10.8.7) and arrives at
\begin{equation}
\Gamma(3d2 \rightarrow 1s) = \frac{2^{23}Gm^3}{3^7
5^{15}(137)^6\hbar^2} = \frac{2^{13} 3^3}{5^{15}} \frac{G}{\hbar}
m^2a^4\omega^5 = 2.5 \times 10^{-44} s^{-1} ,
\end{equation}
where to get the second equality we have used the exact transition
frequency $\omega = 4\hbar/9ma^2$. This gives a lifetime
$\Gamma^{-1} = 4.0 \times 10^{43} s$.

By direct substitution, however, we obtain
\begin{eqnarray}
\Gamma(3d2 \rightarrow 1s)= \Gamma(3d0 \rightarrow 1s)=
\frac{3^8}{5\, 2^{13}} \frac{G}{\hbar}m^2a^4\omega^5 = 4.9 \times
10^{-40} s^{-1},
\end{eqnarray}
which is  about $2.2 \times 10^4$ times larger than Weinberg's and
roughly a factor of ten smaller than the Problem Book's.
Apparently Weinberg's result merely contains a numerical error.

We can nevertheless convince ourselves that the general procedure
is correct, in particular Eq. (\ref{powerquad}), by checking
detailed balance.   We assume
 the spontaneous emission rate given by Eqs. (\ref{Gamma}) and
(\ref{powerquad}).  The rate for stimulated and emission and
absorption follow from (\ref{gr}) and (\ref{me2}).  The crucial
point is that detailed balance will only satisfied when the factor
of $1/2$ in the matrix element (\ref{me}) is replaced by $1/4$
(see Eq. (\ref{HamLIF}) below). Then:
\begin{equation}
\Gamma_{abs} = \Gamma_{se} = \frac{2\pi}{16\hbar}m^2h^2 \omega^4
<|D_{ij}|^2>. \label{Gamabs}
\end{equation}
The intensity $I$ for a gravitational wave is  $I = \omega^2
h^2/8\pi G$. Demanding that the spontaneous plus stimulated
emission rates equal the absorption rate requires
\begin{eqnarray}
e^{-\hbar\omega/kT}\left(G\pi^2 \omega^2 I + G
\omega^5\right)<|D_{ij}|^2> = G\pi^2 \omega^2
<|D_{ij}|^2>,\nonumber
\end{eqnarray}
or
\begin{equation}
I(\omega) = \frac{1}{\pi^2} \frac{\omega^3}{e^{\hbar\omega/kT}
-1}\, , \vspace{2mm}
\end{equation}
 showing that detailed balance is indeed satisfied.
Because the matrix elements cancel out from both sides of this
equation, this procedure cannot provide a  check that they have
been correctly computed
 but it does show
that if Eq. ({\ref{powerquad}) is the correct expression for
spontaneous emission, then $\Gamma_{abs}$ in (\ref{Gamabs}) and
$|<a|H|b>|^2$ in (\ref{me2})
are the correct expression to be used in the calculation of cross sections. \\

The reason for the correction can be found by examining the
Problem Book's field-theoretical calculation.  A proper field
theoretical derivation is the only utterly convincing way to
obtain  the spontaneous emission rate---and the only independent
verification of the semi-classical results.  Why then does the
Problem Book fail to produce the semi-classical answer, which one
would expect it to do?  The crucial step in the authors'
calculation is to write Hamiltonian (\ref{Hamiltonian}) as
\begin{equation}
 H =(8\pi G)^{1/2} \frac{p_ip_j\phi^{ij}}{m}. \label{HamTT}
\end{equation}
Here, $\phi_{\mu\nu} \equiv h_{\mu\nu}/\sqrt G$.  They have also
chosen the same components of the polarization tensor as we did in
\S\ref{matrix}, implying that $T_{\mu\nu} = p_ip_j/m$.

Next, in the standard field-theoretic manner, $\phi$ is decomposed
into a series of harmonic oscillators by the use of creation and
annihilation operators. (The $8\pi G$ is introduced to normalize
the number operator to the energy density; see BR.)

To get $\Gamma$, they compute $<3d|H|1s>$ for the spontaneous
emission of a graviton. This is carried out in by the standard
field-theoretic prescription with the result that in the dipole
approximation
\begin{equation}
\frac {d\Gamma}{d\Omega} = \frac{G\omega}{\pi \hbar m^2}
 |<1s|p_ip_je^{ij}_{k,\lambda}|3d>|^2. \label{dGPB}
\end{equation}
(The conceptual difficulty about using an interaction Hamiltonian
for spontaneous emission is obviated in field theory by saying
that the interaction is with vacuum fluctuations.)

The remainder of the solution consists of a sophisticated
averaging procedure over all spin states via Clebsch-Gordon
coefficients and the Wigner-Eckart theorem.  For a final result
the Problem Book gets $\Gamma = 5.3 \times 10^{-39}s^{-1}$, which
as mentioned is about $10$ times larger than our result.
Nevertheless,  a simple field-theoretic calculation performed with
Eq. (\ref{me2}), does yield exactly the semi-classical answer,
which shows that, not only Weinberg's result, but the Problem
Book's must be incorrect. There are several minor mistakes
involved, which we discuss in BR; here,
however, we do point out a fundamental error that has  been made:\\

In general relativity, coordinate transformations and gauge
transformations are the same thing, and it turns out that the
Hamilonian (\ref{Hamiltonian}) is not gauge invariant.  In setting
$e_{11} = -e_{22} $ and $e_{12} = e_{21}$, the Problem Book
authors have chosen to work in the so-called transverse-traceless,
or TT, gauge, in which the polarization directions are purely
spatial and orthogonal to the direction of the wave's propagation.
It is a textbook exercise to show\cite{Stephani90} that to
first-order in the TT gauge, two particles initially at rest
remain at rest under the passage of a gravitational wave. In other
words, nothing happens, which suggests that the TT gauge cannot
easily be used for this problem. Properly, physics should be
calculated in a locally inertial frame (LIF), in which the effects
of gravity are absent. Otherwise, in the TT gauge for example, not
only are ``non-inertial" forces present, but the laws of
electromagnetism are significantly modified as well.  In a LIF,
such effects are second-order, typically smaller than in the TT
gauge by a factor of $v^2$, where $v$ is the velocity of particles
in the system.

In an LIF, the energy density $\rho$ dominates over the momentum
and pressure, and so the only non-negligible component of the
stress-energy tensor is $T_{00} = \rho$. A straightforward
coordinate transformation from the TT gauge to the LIF (see BR)
yields for the Lagrangian density
\begin{equation}
{\mathcal L}_{LIF} = -\frac{1}{4}\omega^2\rho
he^{TT}_{ij}x^{i}x^{j},
\end{equation}
Thus, the LIF interaction Hamiltonian is
\begin{equation}
H_{LIF} = -\int {\mathcal L} d^3x = \frac{1}{4} \omega^2 m
he^{TT}_{ij}x^{i}x^{j}. \label{HamLIF}
\end{equation}
We see though, in the case $p_i = im\omega x_i$,
 the locally inertial Hamiltonian itself has
exactly the same form as the TT-gauge Hamiltonian in
(\ref{HamTT}), but that each term is smaller by  a factor of \
$\approx$ 2. Thus, it is not entirely surprising  that a TT-gauge
calculation leads to almost the locally inertial result.
Nevertheless, the LIF Hamiltonian (\ref{HamLIF}) is the easiest
one to use in the computation of the cross section. One can carry
out the calculation in the TT gauge (see BR) but in that case one
needs to include the electromagnetic stresses in the stress-energy
tensor.

{\small

\end{document}